\documentclass{emulateapj}

\usepackage{threeparttable,textcomp,multirow}
\usepackage{tensor}
\newcommand{\simgt}{\,\hbox{\lower0.6ex\hbox{$\sim$}\llap{\raise0.6ex\hbox{$>$}}}\,}
\slugcomment{Submitted to ApJ. Lett.}
\shorttitle{A Small Contribution to Reionization from Accretion Shocks}

\shortauthors{Wyithe, Mould \& Loeb}

\begin{document}

\title{The Shocking Truth: \\The small contribution to hydrogen reionization from gravitational infall}

\author{J. Stuart B. Wyithe\altaffilmark{1,3}} 
\author{Jeremy R. Mould\altaffilmark{2,3}} 
\author{Abraham Loeb\altaffilmark{4}} 
\altaffiltext{1}{School of Physics, University of Melbourne,  Parkville, Victoria 3010, Australia}
\altaffiltext{2}{Center for Astrophysics and Super Computing, Swinburne University, Hawthorn Vic 3122, Australia}
\altaffiltext{3}{ARC Centre of Excellence for All-sky Astrophysics (CAASTRO)}
\altaffiltext{4}{Astronomy Department, Harvard University, 60 Garden Street, Cambridge, MA 02138, USA}
\begin{abstract}

It is commonly thought that stars are responsible for reionizing the
Universe. However, deep searches for star-forming galaxies during the
epoch of reionization have not yet found sufficient galaxies to
provide the necessary ionizing flux. Fast accretion shocks associated with
gravitational infall of baryons during the formation of galaxies have
recently been proposed as an alternative method of generating the
required ionising photons. In this {\em Letter} we assess the contribution to
hydrogen reionization from shocked gas associated with gravitational infall. We
find that shocks can ionize no more than a few percent of the cosmic
hydrogen by $z\sim6$. However, the small fraction of ionising
radiation produced by fast accretion shocks would be significantly more
biased than that associated with stars, leading to a modification of
the luminosity weighted source clustering by $\sim10\%$. This modification of the bias may
be measurable with future precision experiments utilising the
redshifted 21cm line to study the distribution of hydrogen during the
reionization era.

\end{abstract}

\keywords{cosmology: theory, reionization, diffuse radiation}

\section{Introduction} 
\label{intro}

Star-bursting galaxies and quasars have been the leading candidates
for the sources of the UV radiation required to reionize the hydrogen
gas in the inter-galactic medium (IGM) \citep{Loeb2010}.  The quasar
population is observed to decline quickly at $z\gtrsim 2.5$
\cite[e.g.,][]{Fan2002} and so it is believed that galaxies
contributed the bulk of UV photons that drove reionization
\citep{Madau1999,Srbinovsky2007,Bolton2007}.  The observed number
counts of high-redshift galaxy candidates~\citep{bouwens2010c,yan2009}
discovered with the Hubble Space Telescope Ultra Deep Field (HUDF)
have been used to build up a statistical description of star-forming
activity at redshift $z\ga7$.
At $z\sim7$, 8.6 and 10.6 the flux limits correspond to absolute
magnitudes $M_{\rm lim}=-18.0$, $-18.3$ and $-18.6$~mag,
respectively. While impressively faint, these observations do not
reach the levels corresponding to the faintest galaxies thought to
exist at these early epochs \cite[e.g.,][]{BL2000}, and the observed
stars in the HUDF are insufficient to reionize the Universe. 

\citet[][]{Trenti2010}
have constructed an empirical model for the luminosity function and
find that the observed population could have reionized the Universe
if it extends to luminosities fainter than observed. In addition to galaxies, the discovery of high redshift gamma ray
bursts \citep[][]{Salvaterra2009,Tanvir2009} can be used to probe
the star formation rate up to $z\sim8$ \citep{Kistler2009}, based on which
\citet[][]{Wyithe2010} found sufficient star-formation to achieve reionization by $z\sim6$.

Thus, there is evidence that stars are capable of reionizing the
Universe. However, until the sources are identified directly, it is
prudent to study alternatives. One such alternative is provided by cooling radiation associated with shock-heated gas, which can process gravitational energy associated with structure formation into an ionising radiation background. For example, \citet[][]{Furlanetto2004} have considered the effect of large scale structure shocks on the IGM during reionization, finding that shocks heat the early IGM and produce a radiation background that effects molecular hydrogen formation and the spin temperature of neutral hydrogen. In a complementary study, \citet[][]{Miniati2004} calculated the UV background from cooling radiation associated with virialised gas that is shock heated during halo formation. They evaluated the ionising rate in the optically thin IGM at $z<6$, and determined that the ionising background at the hydrogen ionisation edge produced following virialisation shocks was likely to be much smaller than for stars, unless supernova feedback could efficiently reheat the galactic gas (in which case the galaxies would likely produce a significant contribution to the ionising flux from star light).  \citet[][]{Miniati2004} concluded that the hard spectrum produced may doubly reionize helium by $z\sim6$, and also described the resulting effects on the thermal history of the IGM. 

Recently, \citet[][]{Dopita2011} presented high resolution simulations of the shocks associated with the supersonic gravitational infall of gas onto a nascent galactic disk that arises when the cooling rate is much shorter than the free-fall time. They argued that this provided a source of photons that could augment the ionising flux from galaxies, and indeed might dominate the reionization of the Universe. If true this finding would have significant implications for studies of reionization. In particular, the direct link between star-formation and the ionisation structure of the IGM would be removed, so that future redshifted 21cm experiments would not provide
a fruitful route to study the first stars. 

In this {\em Letter} we assess the contribution to hydrogen reionization from ionising photons produced in fast radiative accretion shocks.
We note that virialisation \citep[][]{Miniati2004} and fast accretion shocks \citep[][]{Dopita2011} should lead to similar ionizing luminosities since the cooling radiation following the virialization shock contains energy comparable to the gravitational potential energy available to drive the fast accretion shock. In our numerical examples, we adopt the standard
set of cosmological parameters \citep{Komatsu2011}, with values of $\Omega_{\rm b}=0.04$, $\Omega_{\rm m}=0.24$ and $\Omega_\Lambda=0.76$ for
the matter, baryon, and dark energy fractional density respectively,
$h=0.73$, for the dimensionless Hubble constant, and $\sigma_8=0.82$.

\section{Ionizing photons} 

We begin by computing the number of ionizations per baryon processed through shocks. We also discuss the ionisation history due to stars for comparison.

\subsection{Ionizing photons based on collapsed fraction}

\label{ioncol}

\begin{figure*}[htb]
\begin{center}
\includegraphics[width=16cm]{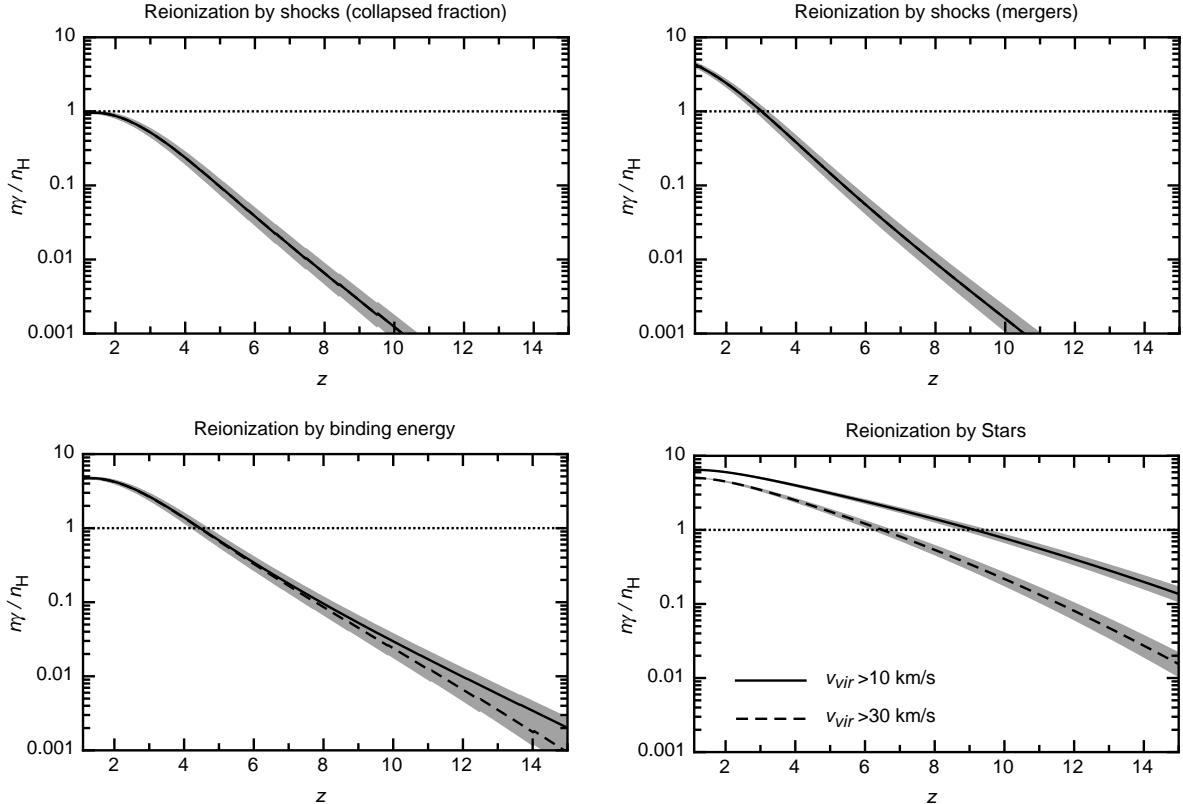}
\caption[short]{Plots of the number of photons produced per hydrogen
atom in the Universe. {\em Upper Left:} The number produced when all
gas that enters a halo shocks once when the halo forms. {\em Upper
Right:} The number produced when gas re-shocks following each merger
(based on the merger rate of halos). {\em Lower Left:} The number
produced assuming all gravitational energy of collapsed gas was
released as photons with 13.6eV. {\em Lower Right:} The number number
of ionizations obtained for stars, assuming $N_\gamma=4000(f_\star
f_{\rm esc})=15$. Results are shown as a function of redshift,
assuming minimum halo masses corresponding to $v_{\rm vir}=10~{\rm
km~s^{-1}}$, and
$v_{\rm vir}=30~{\rm km~s^{-1}}$. The grey strips represent uncertainty in $\sigma_8$.}
\label{plot1}
\end{center}
\end{figure*}

\citet[][]{Dopita2011} presented a fitting formula for the number of ionising photons that enter the IGM per baryon processed through a shock, which in our terminology is
\begin{eqnarray}
\label{Ngamma}
\nonumber
		N_\gamma 	&=& 0.2\left(\frac{v}{100\mbox{km/s}}\right)^2 \hspace{5mm}\mbox{for}\hspace{2mm}100<v<280\,\mbox{km/s}\\
N_\gamma &=& 10\left(\frac{v}{400\mbox{km/s}}\right)^ 5\hspace{5mm}\mbox{for}\hspace{2mm}280<v<400\,\mbox{km/s}\\
\nonumber	N_\gamma 	&=& 10\left(\frac{v}{400\mbox{km/s}}\right)^ 2\hspace{5mm}\mbox{for}\hspace{2mm}v>400\,\mbox{km/s},
\end{eqnarray}
where $v$ is the velocity of the shocked gas. \citet[][]{Dopita2011} argue that $v=\sqrt{2}v_{\rm vir}$ where $v_{\rm vir}$ is the virial velocity of the halo, and we utilise this throughout the current work. The virial velocity is estimated from the halo mass using      
\begin{equation}
\label{rvir}
v_{\rm vir}(M_{\rm halo},z)=  23.4 \left(\frac{M_{\rm
halo}}{10^{8}M_{\odot}h^{-1}}\right)^{\frac{1}{3}}
[\zeta(z)]^\frac{1}{6}\left(\frac{1+z}{10}\right)^\frac{1}{2},
\end{equation}
where $\zeta(z)$ is close to unity and
defined as $\zeta\equiv [(\Omega_m/\Omega_m^z)(\Delta_c/18\pi^2)]$,
$\Omega_m^z \equiv [1+(\Omega_\Lambda/\Omega_m)(1+z)^{-3}]^{-1}$,
$\Delta_c=18\pi^2+82d-39d^2$, and $d=\Omega_m^z-1$ (see equations~22--25 in
Barkana \& Loeb 2001 for more details).

Under the assumption that gas shocks once when a halo forms, the number of photons produced per hydrogen atom in the Universe can be estimated from the \citet[][]{press1974} mass function $dn/dM$ \citep[with modifications due to][]{Sheth1999}
\begin{equation}
\label{fcol}
\frac{n_\gamma}{n_{\rm H}} = \frac{1}{0.76\rho_{\rm b}}\int_{M_{\rm halo}}^\infty dM N_\gamma M_{\rm b} \frac{dn}{dM},
\end{equation}
where $M_{\rm b}=(\Omega_{\rm b}/\Omega_{\rm m})M$ and $\rho_{\rm
b}=(\Omega_{\rm b}/\Omega_{\rm m})\rho_{\rm m}$, are the baryonic mass
inside a dark matter halo and baryonic mass density in the IGM
respectively. Results are shown as a function of redshift in the upper
left panel of Figure~\ref{plot1}, assuming minimum halo masses
corresponding to $v_{\rm vir}=10~{\rm km~s^{-1}}$ (the cooling threshold for hydrogen), and $v_{\rm vir}=30~{\rm
km~s^{-1}}$ \citep[the Jeans threshold in an ionized
IGM, ][]{Dijkstra2004}. The results are independent of this choice owing to the
dominance of massive halos in producing radiation from fast accretion shocks
\citep[][]{Dopita2011}. We find that only a few percent of the IGM is
reionized by $z\sim6$, and $<1\%$ at $z\sim8$, indicating that there
are 1-2 orders of magnitude too few ionising photons produced in
fast accretion shocks to reionize the Universe.

We note that virialisation shocks do not exist in low mass halos due to the presence
of cold flows. Given the halo mass dependent fraction of cold flow
accretion $f_{\rm cold}$ where no shock is produced
\cite[][]{Keres2009,Faucher2010}, one should exclude cold mode accretion
material from the calculation of ionising luminosity associated with virialisation shocks~\citep[][]{Miniati2004}. This would
reduce the predictions in \citet[][]{Miniati2004} by a factor of $(1-f_{\rm
cold})$, where $f_{\rm cold}$ is weighted over the mass and number of
halos. However \citet[][]{Dopita2011} argue that with sufficient resolution the cold flow material is found to shock at the intersection with the nascent galactic disk. As a result no correction for cold flow accretion should be applied to estimates of fast accretion shock produced ionising photons in equation~(\ref{fcol}) or later in this {\em Letter}.

\subsection{ionising photons from stars}

We compare the above result to the number of ionizations obtained for
stars (lower right panel of Figure~\ref{plot1}). Here we utilise
equation~(\ref{fcol}) but assume $N_\gamma=4000(f_\star f_{\rm
esc})=15$, appropriate for a Salpeter IMF with a fiducial value of
$f_\star f_{\rm esc}\sim0.004$, where $f_\star$ and $f_{\rm esc}$ are
the star-formation efficiency and escape fraction of ionising photons
respectively. We again show minimum halo masses
corresponding to $v_{\rm vir}=10~{\rm km~s^{-1}}$ and $v_{\rm
vir}=30~{\rm km~s^{-1}}$, leading to significant variation in the
total number of ionising photons produced. In difference to shock produced photons, these fiducial
stellar populations are easily able to reionize the Universe by
$z\sim6$ (as is well known). This result is plausible since the nuclear efficiency of stars is larger
by many orders of magnitude than the efficiency of converting rest mass to
radiation by a shock, $\sim (v/c)^2 \la 10^{-6}$, for the shock
speeds of interest ($v\la 300~{\rm km~s^{-1}}$). 

\subsection{Ionizing photons based on merger rates}

The calculation in \S~\ref{ioncol} utilises each baryon only once, whereas, unlike the case for stars, a baryon may be processed through shocks several times during the hierarchical formation of a galaxy. We therefore recalculate the contribution to reionization from fast accretion shocks based on the merger rate of halos. Specifically, when a halo of mass $M_2<M_1$ merges with a halo of mass $M_1$, we assume that the baryons contained within halo 1 are shocked at $v=\sqrt{2}v_{vir}(M_1+M_2)$. We assume that the star-formation efficiency is negligible so that all baryons are available to shock during each merger. The resulting expression for the rate of production of all baryons in galaxies is 
\begin{eqnarray}
\label{merge}
\nonumber
&&\hspace{-7mm}\frac{d}{dz}\left(\frac{n_\gamma}{n_{\rm H}}\right) = \frac{1}{0.76\rho_{\rm b}}\int_{M_{\rm halo}}^\infty dM_{1} \frac{dn}{dM_1}\\
&&\hspace{5mm} \int_{M_{\rm halo}}^{M_1}  dM_2 \,N_\gamma \times (M_{\rm b,1}+M_{\rm b,2}) \left.\frac{d^2N}{dz dM_2}\right|_{M_1},
\end{eqnarray}
yielding
\begin{equation}
\frac{n_\gamma}{n_{\rm H}} = \int_\infty^z dz^\prime \frac{d}{dz}\left(\frac{n_\gamma}{n_{\rm H}}\right).
\end{equation}
Here $\left.\frac{d^2N}{dz dM_2}\right|_{M_1}$ is the number of mergers of a halo with mass between $M_2$ and $M_2 + dM_2$ that merge with a halo of mass $M_1$ in a redshift interval $dz$. 

The assumption that baryons shock during each merger is
optimistic. While mergers will certainly increase the virial energy
per baryon, it is not clear how the baryons would be recycled through
virialization shocks on multiple occasions if they cool efficiently to
produce the ionizing radiation. If the baryons cool from the halo into
a disk, they will not be heated to the virial temperature of the halo
(unless they have been expelled through some feedback process) and so
should not participate in the halo shock of the next merger. Thus, our
results from equation~(\ref{merge}) represent the maximum ionisation
flux possible from the fast accretion shock mechanism summarised in
equation~(\ref{Ngamma}).

Results are shown as a function of redshift in the upper right panel
of Figure~\ref{plot1}, assuming minimum halo masses corresponding to
$v_{\rm vir}=10~{\rm km~s^{-1}}$ and $v_{\rm vir}=30~{\rm km~s^{-1}}$ (again with no
discernible difference owing to the dominance of massive halos). We
find that the re-processing of baryons through shocks in multiple
mergers increases the ionising photon output by a factor of a few
relative to the collapsed fraction calculation. However, we still find
that only $\sim5$ percent of the IGM is reionized by $z\sim6$, and
$\sim1\%$ at $z\sim8$, indicating that shocks provide insufficient
ionising luminosity to reionize the IGM.

The difference between our findings and the results
presented in \citet[][]{Dopita2011} originates partly from
the adoption of a large value of $\sigma_8=0.9$ in that work, and partly
from an error in the calculation method (Lawrence Krauss, private
communication). \citet[][]{Dopita2011} calculate the accretion rate
and corresponding ionising luminosity as a function of halo mass (see
their Figure 2). They then integrate over the mass-function and
redshift. This procedure effectively sets the rate at which gas in the
halo doubles to be equal to the inverse of dynamical time
at the virial radius (which is shorter by an order of magnitude than
the Hubble time), and so does not
account for the duty-cycle of the shocks (which should be only
$\sim0.1$). As a result, the calculation in \citet[][]{Dopita2011}
accretes an order of magnitude more gas than available per halo.

\subsection{A maximum of ionizing photons based on binding energy}

In the previous subsections we have estimated the number of ionising photons available per hydrogen in the IGM. We next estimate the maximum number of photons that equal the binding energy of all baryons in the halo. This should provide an upper limit to the number of ionising photons produced by shocks \citep[in the absence of significant feedback,][]{Miniati2004}, and hence an upper limit on the contribution of shocks to reionization. For this calculation, we again appeal to equation~(\ref{fcol}), setting $N_\gamma$ to be
\begin{equation}
\label{binding}
N_\gamma = \frac{{1\over 2}M_{\rm b} v_{\rm vir}^2}{13.6 \mbox{eV}}.
\end{equation}
Results are shown in the lower-left panel of Figure~\ref{plot1}
assuming minimum halo masses corresponding to $v_{\rm vir}=10~{\rm
km~s^{-1}}$ and $v_{\rm vir}=30~{\rm km~s^{-1}}$. We find that the
total gravitational energy available for ionisation of hydrogen
corresponds to less than 1 ionising photon per 3 hydrogens by $z\sim6$
and less than 1 ionising photon per 10 hydrogens by $z\sim8$.

\section{Reionization Histories}

Next we use the estimate of flux based on our merger calculation of
ionising radiation from shocks as the source term in a calculation of
the reionization history. \citet[][]{Miralda2000} presented a model
which allows the calculation of an effective recombination rate in an
inhomogeneous universe by assuming a maximum overdensity ($\Delta_{\rm
c}$) penetrated by ionizing photons within HII regions. The model
assumes that reionization progresses rapidly through islands of lower
density prior to the overlap of individual cosmological ionized
regions. Following the overlap epoch, the remaining regions of high
density are gradually ionized. \citet[][]{Wyithe2003} employed this
prescription within a semi-analytic model of reionization,
and we refer the reader to that paper for details of the model. 
Within this formalism, the epoch of overlap is precisely defined as the time when the volume fraction $Q$ of the universe ionized up to an overdensity $\Delta_{\rm c}$, reaches unity.  After the overlap epoch, ionizing photons will experience attenuation due to residual overdense pockets of HI gas. The model also follows the mass averaged ionized fraction ($Q_{\rm m}$).

\begin{figure}[htb]
\begin{center}
\includegraphics[width=8cm]{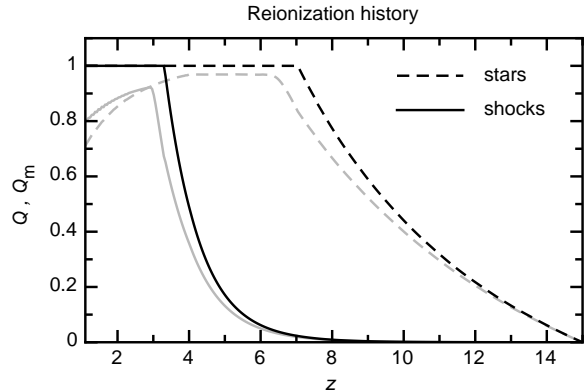}
\caption[short]{Plots of the reionization history of the IGM and the subsequent post-overlap evolution due to ionising sources from merger induced shocks (solid lines), and stars with $N_\gamma=15$ (dashed lines). The case shown corresponds to a value for the critical overdensity prior to the overlap epoch of $\Delta_{\rm c}=5$, and both the volume averaged (thick lines) and mass-averaged (thin lines) ionisation fractions are shown. }
\label{plot2}
\end{center}
\end{figure}

Figure~\ref{plot2} shows the resulting model for the reionization of the IGM and the subsequent post-overlap evolution due to ionising sources from fast accretion shocks (solid lines). Here we have used an ionising photon rate based on equation~(\ref{merge}). The case shown corresponds to a value for the critical overdensity prior to the overlap epoch of $\Delta_{\rm c}=5$, and both the volume averaged (dark lines) and mass-averaged (grey lines) ionisation fractions are shown. We find that shocks can reionize less than 10\% of the IGM (by volume or mass) prior to $z\sim6$, and cannot complete reionization until $z\sim3$. For comparison we compute the reionization history for stars (dashed lines), where the ionising photon production rate is based on $d(n_\gamma /n_{\rm H})/dz$, with $n_\gamma /n_{\rm H}$ based on equation~(\ref{fcol}) with $N_\gamma=15$. In this model, stars complete reionization by $z\sim8$, at which time the relative contributions from stars and fast accretion shocks differ by a factor in excess of 100.

\section{implications for the bias of ionising sources and 21cm studies}

In this {\em Letter} we have demonstrated that recent estimates of the
ionising luminosity from fast accretion shocks associated with galaxy
formation are not sufficient to drive reionization. However the ionising photons produced
by shocks are dominated by massive halos \citep[][]{Miniati2004,Dopita2011}. This is in contrast to the
ionising radiation from stars, which is both predicted and observed to
be dominated by low mass galaxies. As a result, the ionising radiation
produced in shocks is significantly more biased relative to the
underlying large scale density of the IGM than are ionising photons
produced in galaxies. It is easy to see the physics of the dominance of massive halos by noting that the collapse energy available in equation~(\ref{binding}) is proportional to $v_{\rm vir}^6$ (or $M_{\rm halo}^{2}$), whereas the stellar mass (assuming a constant mass-to-light ratio) is proportional to $v_{\rm vir}^3$ (or $M_{\rm halo}$).

\begin{figure}[t]
\begin{center}
\includegraphics[width=8cm]{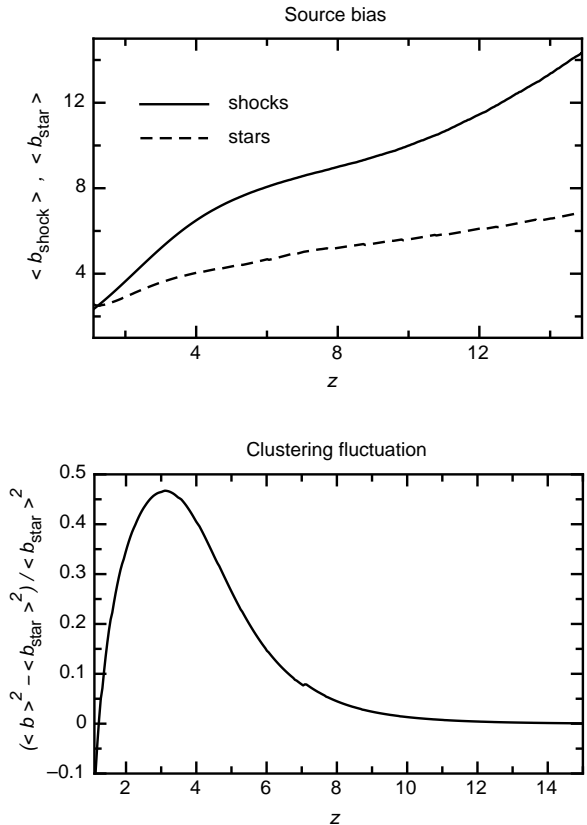}
\caption[short]{Plots of the luminosity weighted galaxy bias. {\em Upper Panel:} the bias is plotted as a function of redshift due to ionising sources from merger induced shocks (solid lines), and stars with $N_\gamma=15$ (dashed lines). {\em Lower Panel:} the fractional change in observed clustering of ionising radiation when compared with the stars-only history. }
\label{plot3}
\end{center}
\end{figure}

The ionisation structure of the IGM, particularly the scale of HII
regions produced is a sensitive function of the bias of ionising
sources \citep[][]{McQuinn2007}. It is this relation between the bias
of ionising sources and the resulting ionisation structure during
reionization that motivates redshifted 21cm experiments with the
ultimate aim of connecting galaxy properties to the power-spectrum of
21cm fluctuations \citep{Barkana2009}. Here we quantify
the effect of fast accretion shocks on the bias of ionising sources.  The halo bias
$b$ for a halo mass $M$ at redshift $z$ may be approximated using the
\citet[][]{press1974} formalism, modified to include non-spherical
collapse \citep[][]{Sheth2001}. The power-spectrum of the space
distribution of sources is proportional to $b$ squared. The luminosity
weighted bias of ionising radiation produced by shocks arising in
mergers can be evaluated using the expression
\begin{eqnarray}
\label{bmerge}
\nonumber
&&\hspace{-5mm}\langle b_{\rm shock} \rangle=\left[\frac{d}{dz}\left(\frac{n_\gamma}{n_{\rm H}}\right)\right]^{-1} \frac{1}{0.76\rho_{\rm b}}\int_{M_{\rm halo}}^\infty dM_1 \frac{dn}{dM_1}\\
&&\hspace{5mm} \int_{M_{\rm halo}}^{M_1}  dM_2\, N_\gamma \times (M_{\rm b,1}+M_{\rm b,2}) \left.\frac{d^2N}{dz dM_2}\right|_{M_1} \, b,
\end{eqnarray}
where the bias $b$ is evaluated at a mass $M_1+M_2$. The resulting bias is plotted as a function of redshift in the upper panel of Figure~\ref{plot3}. Prior to reionization, fast accretion shock powered ionising sources have a luminosity weighted bias of $\langle b_{\rm shock}\rangle\sim10$.

For comparison, we calculate the luminosity weighted bias for stellar
sources ($\langle b_{\rm star}\rangle$) based on the derivative of
equation~(\ref{fcol}), and in analogy with equation~(\ref{merge}). The
result is also plotted in the upper panel of Figure~\ref{plot3}. Prior
to reionization, stellar ionising sources have a smaller luminosity
weighted bias of $\langle b_{\rm star}\rangle\sim4$. Finally, we
evaluate the luminosity weighted bias $\langle b\rangle$ obtained
when fast accretion shock powered ionisation sources are added to the stellar sources
needed for reionization. We then calculate the fractional change in
observed clustering of ionising radiation relative to the stars-only
reionization history [$(\langle b\rangle^2-\langle b_{\rm
star}\rangle^2)/\langle b_{\rm star}\rangle^2$]. This fractional
change is plotted in the lower panel of Figure~\ref{plot3}, which
shows that the power-spectrum of ionising sources is increased by
$\sim 10\%$ owing to ionising radiation produced in shocks. This
change in the clustering of ionising sources will lead to comparable
changes in the amplitude of redshifted 21cm fluctuations
\citep[][]{Wyithe2007}, that will be detectable by planned low
frequency radio telescopes \citep[][]{WyitheW2009}.

\section{summary}
Based on recent high resolution simulations \citep[][]{Dopita2011}, we
have quantified the contribution that gravitationally powered fast accretion shocks
during galaxy formation can make to the reionzation of
hydrogen. We find that ionising
radiation from fast accretion shocks represents a negligible contribution to the
overall reionization history of hydrogen, leaving the dominant contribution to be
provided by stars. This conclusion is consistent with expectations based on observations of cosmic background radiation. The energy released by star formation at high redshift is stored in the cosmic infrared background with $\nu L_\nu \sim 10$ nW/m$^2$/Str \citep[][]{Hauser2001}. On the other hand, any energy surplus from gravitational shocks is stored in the cosmic soft x-ray background at $\sim0.01$ nW/m$^2$/Str. This three orders of magnitude difference is suggestive of the relative efficiency of shocks and star formation in illuminating the Universe at high redshift.

Despite their small contribution to hydrogen reionization, shocks may have observable consequences for studies of the reionization era. As discussed by \citet[][]{Miniati2004}, the harder spectrum associated with shocks 
will lead to a modification of the thermal history. In particular, the reionization of hydrogen by shocks would be accompanied by reionization of singly ionized Helium, thus heating the IGM to levels above those observed at $z\sim5$ \cite[][]{Becker2011}. While this likely rules out reionization by shocks, independently from the hydrogen ionisation photon budget, heating by shocks may still have important consequences for star-formation at high redshift \citep[e.g.][]{Dijkstra2004}. 

In addition, we find that because the small contribution from fast accretion shocks is produced in highly biased galaxies, their
presence modifies the mean clustering bias of the combined ionising
radiation. This modification will likely lead to observable changes in
the redshifted 21cm fluctuations from neutral hydrogen during
reionization, and so will need to be considered in analyses which aim
to use precision measurements of 21cm fluctuations to study the
properties of very high redshift galaxies.

\vspace{5mm}

{\bf Acknowledgments} We thank James Bolton, Jonathan Bittner and Lawrence Krauss
for helpful discussions during this work.  The Centre for All-sky
Astrophysics is an Australian Research Council Centre of Excellence,
funded by grant CE11E0090.  AL was supported in part by NSF grant
AST-0907890 and NASA grants NNX08AL43G and NNA09DB30A.

\vspace{1mm}



\end{document}